\documentclass[]{MSword}

\usepackage[english]{babel}
\usepackage{csquotes}
\usepackage{lipsum}
\usepackage{caption}
\usepackage{siunitx}
\usepackage{pifont}
\usepackage{graphicx}
\usepackage{adjustbox}
\usepackage{amsmath}
\usepackage{upgreek}
\usepackage{authblk}
 
\title{Tailoring coherent microwave emission from a solid-state hybrid system for room-temperature microwave quantum electronics}
\author[1$\dagger$]{Kaipu Wang}
\author[1$\dagger$*]{Hao Wu}
\author[1]{Bo Zhang}
\author[1]{Xuri Yao}
\author[2]{Jiakai Zhang}
\author[3]{Mark Oxborrow}
\author[1*]{Qing Zhao}
\affil[1]{Center for Quantum Technology Research and Key Laboratory of Advanced Optoelectronic Quantum Architecture and Measurements (MOE), School of Physics, Beijing Institute of Technology, Beijing 100081, China}
\affil[2]{Xi'an Electronic Engineering Research Institute, Xi'an 710100, China}
\affil[3]{Department of Materials, Imperial College London, South Kensington, London SW7 2AZ, UK}
\affil[$\dagger$]{These authors contributed equally to this work.}
\affil[*]{Corresponding author. Email: hao.wu@bit.edu.cn, qzhaoyuping@bit.edu.cn}
\date{}

\begin{document}
\captionsetup[figure]{labelfont={bf},labelformat={default},labelsep=period,name={Fig.}}
\captionsetup[figure]{font={stretch=1.2}}
\maketitle

\begin{center}
\section*{Abstract}
\end{center}
\textbf{Quantum electronics operating in the microwave domain are burgeoning and becoming essential building blocks of quantum computers, sensors and communication devices. However, the field of microwave quantum electronics has long been dominated by the need for cryogenic conditions to maintain the delicate quantum characteristics. Here we report on a solid-state hybrid system, constituted by a photo-excited pentacene triplet spin ensemble coupled to a dielectric resonator, that is for the first time capable of both coherent microwave quantum amplification and oscillation at X band via the masing process at room temperature. By incorporating external driving and active dissipation control into the hybrid system, we achieve efficient tuning of the maser emission characteristics at around 9.4 GHz, which is key to optimizing the performance of the maser device. Our work not only pushes the boundaries of the operating frequency and functionality of the existing pentacene masers, but also demonstrate a universal route for controlling the masing process at room temperature, highlighting opportunities for optimizing emerging solid-state masers for quantum information processing and communication.}  
\section*{Introduction}
\hspace{2em}In quantum information processing and communication (QIPC) systems, microwave amplifiers and oscillators are vital devices for qubits manipulation\cite{bardin2020quantum}, quantum-state readout\cite{heinsoo2018rapid,walter2017rapid}, signal transmitting\cite{magnard2020microwave,assouly2023quantum} and amplification\cite{peng2021cryogenic,bardin2021cryogenic,patra2017cryo}. Realization of ultra-low-noise features (specifically in terms of the noise figure and phase noise) of such electronics is of major importance for securing the accuracy, sensitivity and efficiency of QIPC systems. To this end, quantum mechanical approaches to coherent microwave emission have been extensively studied for pursuing the quantum amplifiers\cite{sherman2022diamond,massel2011microwave,yuimaru2023ultra,koppenhofer2022dissipative} and oscillators\cite{bourgeois2005maser,yan2021low,liu2015injection} with quantum-limited noise performance\cite{schawlow1958infrared,withington2022quantum} superior to the classical devices. Among the approaches, microwave superradiance\cite{dicke1954coherence} and masing\cite{gordon1955maser}, as "Drosophila" in the field of quantum electronics first proposed in 1950s, have yet to fade away. The former is a collective \textit{spontaneous} emission phenomenon\cite{gross1982superradiance} resulting from the spontaneous phase-locking of a collection of emissive dipoles with transition frequency in the microwave domain, whereas the latter originates from the \textit{stimulated} emission of radiation\cite{einstein1916quantentheorie} in the dipole system. The exploitation of either phenomenon into solid-state platforms is a straightforward route for realizing the aforementioned microwave quantum electronics, however, few attempts have been undertaken at room temperature due to the substantially increased spin-lattice/spin-spin relaxation rates of the dipoles in the microwave gain media\cite{gill1962spin,reynhardt1998temperature} and the operating requirements of the superconducting microwave elements, e.g. resonators\cite{gao2008physics} and Josephson junctions\cite{kautz1981ac}.

The remedy to the above problem is to hybridize the quantum modules whose intrinsic properties are suitable for room-temperature applications. Tremendous strides have been made in the past decade in developing the solid-state electronic spin systems with long spin-lattice relaxation and coherence times at room temperature. The prominent systems are the negatively charged nitrogen-vacancy centers in diamond (NV$^-$ diamond)\cite{doherty2013nitrogen}, spin defects in silicon carbide (SiC)\cite{koehl2011room,riedel2012resonant}, boron vacancies in hexagonal boron nitride (hBN)\cite{gottscholl2021room} and pentacene triplets in \textit{p}-terphenyl single crystals\cite{wu2019unraveling}. Complementing this, dielectric materials have become strong candidates for the fabrication of the microwave resonators with high quality factors ($Q$)\cite{breeze2007enhanced,breeze2011better}, small mode volumes ($V_\textrm{mode}$)\cite{breeze2015enhanced} and uniform confined electromagnetic fields\cite{eisenach2018broadband} facilitating efficient control and readout of quantum states at ambient conditions\cite{ebel2021dispersive,eisenach2021cavity}. The explored natural advantages of the complementary quantum systems, i.e. the solid-state spin systems and dielectric microwave resonators, have laid the foundation for achieving microwave superradiance and masing at room temperature. To date, the feasibility of the room-temperature microwave superradiance has been theoretically verified in the solid-state hybrid quantum systems comprised of the NV centers\cite{wu2022superradiantNV,zhang2022cavity} or the pentacene triplets\cite{wu2022superradiant} coupled to dielectric resonators while the experimental investigations were still in the cryogenic regime\cite{angerer2018superradiant,gottscholl2022superradiance,kersten2023triggered}. In contrast, the room-temperature solid-state maser technology utilizing the similar hybrid quantum systems has undergone rapid growth over the last decade\cite{oxborrow2012room,kraus2014room,jin2015proposal,breeze2018continuous,wu2020room,arroo2021perspective,ng2023move,attwood2023n}. The increasing maser prototypes have been experimentally demonstrated to reveal their potential as room-temperature microwave quantum electronics for magnetic-field sensing\cite{wu2022enhanced} and tunable coherent microwave generation\cite{zollitsch2023maser}.

The triplet electron spins of the photo-excited pentacene molecules doped in \textit{p}-terphenyl are one of the mature solid-state spin systems that can be employed as a room-temperature maser gain medium. Compared with the other popular candidate NV diamond, such organic mixed spin systems possess substantially higher spin densities\cite{miyanishi2021room} enabling the more powerful maser emission\cite{arroo2021perspective}. Additionally, the ease of chemical modifications\cite{bogatko2016molecular} and bulk preparation\cite{cui2020growth} offers the advantages of tailorable functionalities, reproducibility and low cost for practical maser devices. Nonetheless, the pentacene masers demonstrated so far are all microwave oscillators\cite{oxborrow2012room,breeze2015enhanced,salvadori2017nanosecond,wu2020invasive,wu2020room,ng2023maser} with solely a single operating frequency in the L-band ($\sim1.45$ GHz) which severely restricts their applications. Moreover, controllable maser emission is of great importance for reaching the optimal performance of the device, however, such capacity is still unexplored.

In this Article, we report for the first time a dual-function pentacene maser that exhibits the capabilities of quantum amplification and oscillation in the X-band ($\sim9.4$ GHz) at room temperature. The device harnesses the orientation-dependent Zeeman effect in a solid-state hybrid quantum system that comprises of optically polarized triplet spins of pentacene embedded in \textit{p}-terphenyl and a sapphire resonator. The parameter space of the hybrid system including the spin decoherence time $T_\textrm{2}$ and the resonator's conversion factor $\mathit{\Lambda}$ can be characterized \textit{in situ} via the transient electron paramagnetic resonance (trEPR) technique. We demonstrate the ability to control the gain characteristics of the maser-based quantum amplifier with the external microwave driving and systematically calibrate the amplifier performance. For the maser-based quantum oscillator, we utilize the active control of the power dissipation in the hybrid system for maser-threshold tuning and reveal a linear correlation between the reciprocal loaded quality factor $1/Q_\textrm{L}$ of the resonator and the maser threshold. Our work not only pushes the boundaries of the operating frequency and functionality of the existing pentacene masers, but also provides a universal approach for optimizing the performance of the maser-based quantum electronics, thus facilitating their applications in QIPC systems.

\section*{Results}

\begin{figure}[htbp!]
    \centering
    \includegraphics[width=\textwidth]{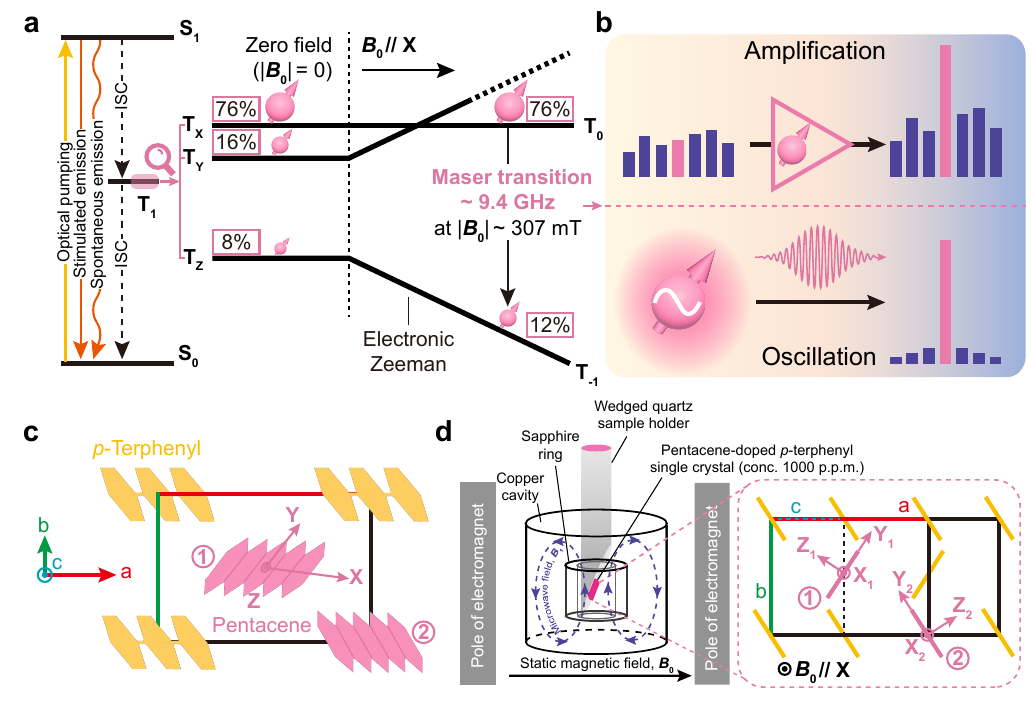}
    \caption{\textbf{Functionality and architecture of the solid-state hybrid system.} \textbf{a,} Room-temperature maser mechanism under an external magnetic field. Upon optical pumping, the electron spins of pentacene are promoted from the singlet ground state S$_{0}$ to the excited singlet state S$_{1}$ and subsequently undergo the transitions back to S$_{0}$ via stimulated and spontaneous emission as well as the intersystem crossing (ISC). Due to the ISC process, the lowest metastable triplet state, T$_{1}$ of pentacene becomes populated in its non-degenerate sublevels, T$_\textrm{X}$, T$_\textrm{Y}$ and T$_\textrm{Z}$ with a population ratio of 76\%:16\%:8\% at zero field. Applying a static magnetic field (|\textbf{\textit{B}}$_\textbf{0}|\sim307$ mT) parallel to the pentacene molecular X axes splits the zero-field triplet sublevels T$_\textrm{Y}$ and T$_\textrm{Z}$ through the Zeeman interaction and the corresponding high-field triplet sublevels are denoted by T$_\textrm{+1}$ (not shown) and T$_\textrm{-1}$, respectively. In the high-field limit,T$_\textrm{X}$ is denoted by T$_\textrm{0}$ when \textbf{\textit{B}}$_\textbf{0}$//X and its energy as well as the triplet spin population stays the same as at zero field, whereas the remaining spin populations are equally redistributed in T$_\textrm{+1}$ and T$_\textrm{-1}$, producing the population inversion between T$_\textrm{0}$ and T$_\textrm{-1}$ for the maser transition at $\sim9.4$ GHz. \textbf{b,} Schematic illustration of the maser emission applied in microwave quantum electronics: room-temperature microwave quantum amplifiers (top) and oscillators (bottom). \textbf{c,} Room-temperature crystal structure of pentacene-doped \textit{p}-terphenyl with the crystal axes labelled. Two possible sites of pentacene molecules (pink) substitutionally doped in the lattice of \textit{p}-terphenyl (yellow) are denoted by \ding{172} and \ding{173}. X and Y, the in-plane molecular symmetry axes of pentacene; Z, the out-of-plane molecular symmetry axis. \textbf{d,} Left: main components of the solid-state hybrid system. A pentacene-doped \textit{p}-terphenyl single crystal (pink) mounted on a wedged sample holder is placed within the bore of a sapphire resonator housed inside a copper cavity. Purple dotted ellipses, the microwave magnetic field (\textbf{\textit{B}}$_\textbf{1}$) flux of the TE$_{01\updelta}$ mode of the sapphire resonator. Right: the sample holder is horizontally rotatable for achieving the optimal condition that the X axes of the pentacene molecules in the two inequivalent sites can be simultaneously aligned with the orientation of the static magnetic field \textbf{\textit{B}}$_\textbf{0}$ generated by an electromagnet. X$_\textrm{m}$, Y$_\textrm{m}$ and Z$_\textrm{m}$ (m = 1 and 2), the molecular symmetry axes of the pentacene molecules in sites \ding{172} and \ding{173}. Pink and yellow bars, pentacene and \textit{p}-terphenyl molecules viewed along the molecular X axes.}
    \label{fig:concept}
\end{figure}

\hspace{2em} \textbf{Principle and design of the device}

The solid-state hybrid quantum system used in the device is the photo-excited triplet spins of pentacene molecules doping crystalline \textit{p}-terphenyl coupled to the axisymmetric transverse electric (TE$_{01\updelta}$) mode of a microwave resonator. The X-band maser emission from the hybrid system is achieved by exploiting the triplet mechanism\cite{oxborrow2012room} in the presence of a static magnetic field \textbf{\textit{B}}$_\textbf{0}$, as shown in Fig.~\ref{fig:concept}a. At room temperature, the optical pumping at 590 nm can excite the pentacene electron spins from the singlet ground state S$_{0}$ to the singlet excited state S$_{1}$. Subsequently, the promoted spins will decay back to S$_{0}$ via the combined processes of the optical stimulated and spontaneous emission\cite{wu2023toward}, as well as the intersystem crossing (ISC). The intriguing property of the pentacene's ISC process is that the metastable triplet state T$_{1}$ formed during the ISC is non-degenerate at zero field where the three sublevels T$_\textrm{X}$, T$_\textrm{Y}$ and T$_\textrm{Z}$ are anisotropically populated with a ratio of $P_\textrm{X}:P_\textrm{Y}:P_\textrm{Z}=$76\%:16\%:8\%\cite{sloop1981electron} and conserve the strong population inversion (prerequisite for masing) upon their instantaneous formation. The transition frequencies among the zero-field triplet sublevels are determined by the zero-field-splitting parameters $D=1395.57$ MHz and $E=-53.35$ MHz\cite{yang2000zero} which give rise to the reported L-band maser transition between T$_\textrm{X}$ and T$_\textrm{Z}$ with a frequency equal to $D+|E|\sim1.45$ GHz\cite{oxborrow2012room}. 

To achieve the higher maser transition frequency, \textbf{\textit{B}}$_\textbf{0}$ is applied to increase the energy splittings among the triplet sublevels via the electronic Zeeman effect. When \textbf{\textit{B}}$_\textbf{0}$ is aligned with one of the pentacene's molecular axes shown in Fig.~\ref{fig:concept}c, the energy of the corresponding sublevel stays identical to that at zero field, while the splittings between the remaining two sublevels is increased as a function of |\textbf{\textit{B}}$_\textbf{0}|$. In the high-field approximation, the unchanged triplet sublevel is denoted as T$_\textrm{0}$ and another two, according to their energetic ordering, are denoted as T$_\textrm{+1}$ and T$_\textrm{-1}$, respectively. In the experiment, we align \textbf{\textit{B}}$_\textbf{0}$ with the in-plane long axis (i.e. X-axis in Fig.~\ref{fig:concept}c) of the pentacene molecules for achieving the highest spin polarization between T$_\textrm{0}$ and T$_\textrm{-1}$ ($P_\textrm{0}:P_\textrm{-1}=76\%:12\%$) since the populations in the high-field sublevels obey $P_\textrm{0}=P_\textrm{X}$, $P_{\pm1}=\frac{1}{2}(P_\textrm{Y}+P_\textrm{Z})$ . Analogously, the rule also holds for the remaining two canonical \textbf{\textit{B}}$_\textbf{0}$ alignments\cite{richert2017delocalisation}. When pentacene molecules are doped in the lattice of \textit{p}-terphenyl at room temperature, there are two inequivalent doping sites labelled as \ding{172} and \ding{173} in Fig.~\ref{fig:concept}c, where the X axes of the two groups of pentacene molecules are parallel to each other while their short in-plane (Y) axes form an angle of \SI{60}{\degree}\cite{lang2006mapping} (Fig.~\ref{fig:concept}d). Therefore, aligning \textbf{\textit{B}}$_\textbf{0}$ with the common X-axis can eliminate the difference between the two groups due to their spectral identity that will be beneficial for making full use of the doped pentacene molecules for masing. Based on the spin Hamiltonian of the pentacene triplet spins ({Methods}), we can obtain the spin transition frequency $\omega_\textrm{s}$ between T$_\textrm{0}$ and T$_\textrm{-1}$ that is about $2\pi\times9.4$ GHz when \textbf{\textit{B}}$_\textbf{0}$//X with a field strength of 307 mT. If the inverted two-level system constructed by T$_\textrm{0}$ and T$_\textrm{-1}$ is resonant with the resonator mode $\omega_\textrm{c}$, the stimulated emission of microwave photons, i.e. masing, will occur that can be exploited for either quantum amplification or oscillation (Fig.~\ref{fig:concept}b) depending on the relationship between the rate of maser emission $\kappa_\textrm{m}$ and the specific loss rates associated with the resonator: (i) the rate of intrinsic loss $\kappa_\textrm{0}$; (ii) the rate of external coupling loss $\kappa_\textrm{e}$. The maser can be configured as an amplifier if $\kappa_\textrm{0}<\kappa_\textrm{m}<\kappa_\textrm{0}+\kappa_\textrm{e}$. If $\kappa_\textrm{m}>\kappa_\textrm{0}+\kappa_\textrm{e}$, the maser can work as an oscillator\cite{breeze2015enhanced}. 

Fig.~\ref{fig:concept}d shows the schematic setup of the device. A pentacene-doped \textit{p}-terphenyl single crystal with a doping concentration of 1000 p.p.m. is mounted on the wedged surface of a quartz sample holder which is horizontally rotatable for aligning the pentacene X-axes with the static magnetic field \textbf{\textit{B}}$_\textbf{0}$ generated by an electromagnet. The mounted sample is loaded inside the bore of a microwave resonator which is a sapphire ring supporting the TE$_{01\updelta}$ mode with an unloaded quality factor $Q_\textrm{0}$ $=2.2\times10^4$ and a mode volume $V_\textrm{mode} = 0.22$ cm$^{3}$ ({Methods}). Due to the characteristic of the TE$_{01\updelta}$ mode, the microwave field \textbf{\textit{B}}$_\textbf{1}$ driving the spin transition is orthogonal to \textbf{\textit{B}}$_\textbf{0}$. The sapphire resonator is housed inside a copper cavity to suppress the microwave radiative loss. A hole is made on the cavity side wall for the pulsed optical pumping of the sample. More details of the setup can be found in {Methods}.

\textbf{Optimization of the crystal orientation in the device}
\begin{figure}[htbp!]
    \centering
    \includegraphics[width=\textwidth]{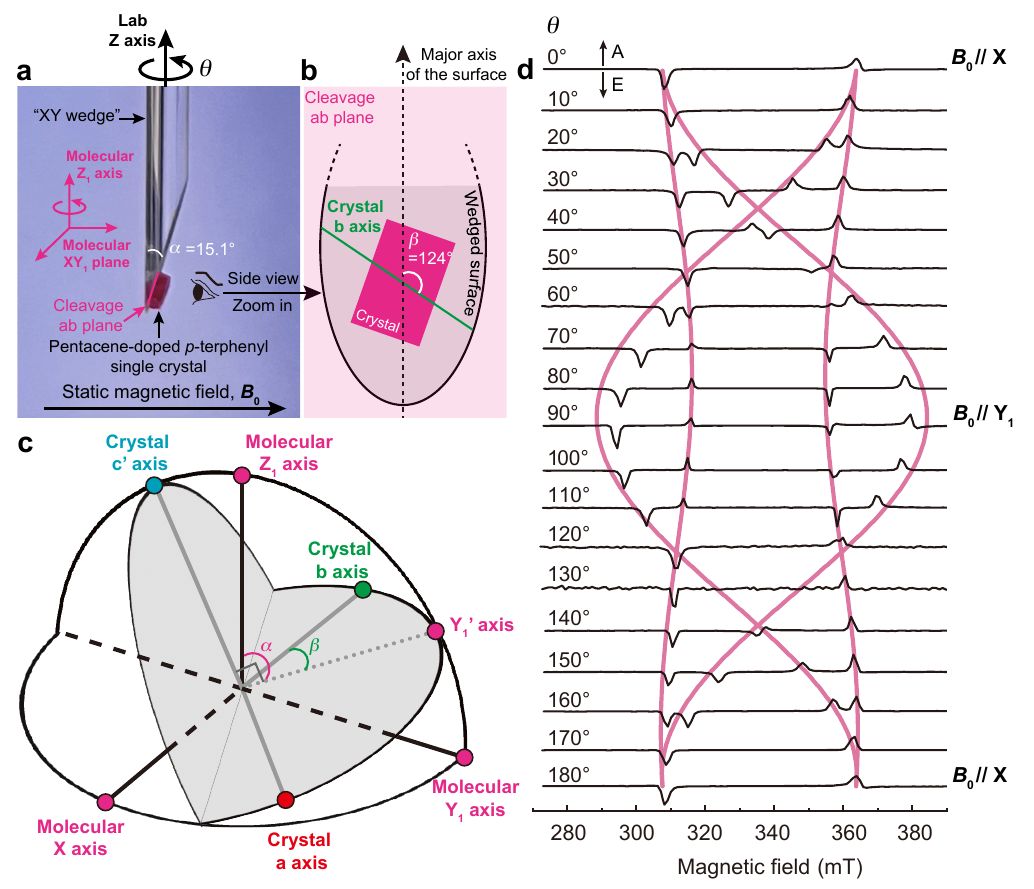}
    \caption{\textbf{Angular-dependent trEPR spectra of the pentacene triplet spins}. \textbf{a,} Photo of the mounted sample. The cleavage ab plane of the pentacene-doped \textit{p}-terphenyl single crystal is mounted on the wedged surface of the quartz sample holder ("XY wedge") to ensure the molecular XY$_1$ planes (\textbf{\textit{B}}$_\textbf{0}$ within them) of a group of the pentacene molecules (e.g. \ding{172} in \textbf{Fig. 1c}) are horizontal and their molecular Z axes are aligned with the sample holder's cylindrical axis, i.e. the lab Z-axis. The sample holder is horizontally rotated along the lab Z-axis with $\theta$ denoting the angle departure from the condition of \textbf{\textit{B}}$_\textbf{0}$//X. $\alpha$, angle of the wedge. \textbf{b,} Schematic of the sample mounting condition on the wedged surface. $\beta$, angle between the major axis of the wedged elliptical surface and the crystal's b-axis. \textbf{c,} Microscopic origins of $\alpha$ and $\beta$. Both angles are determined according to the relationship between the crystal (grey lines) and molecular (black lines) coordinate systems. The projection of the molecular Y$_1$-axis on the crystal ab plane is marked as Y$_1$'. $\alpha$ and $\beta$ are the angles of the molecular Z$_1$-axis and crystal b-axis relative to Y$_1$'. \textbf{d,} trEPR spectra of pentacene-doped \textit{p}-terphenyl measured at different $\theta$ values. The spectra (black) at different orientations are plotted with the measured trEPR amplitudes as a function of the magnetic-field strengths. Pink traces, simulations of the rotation patterns. A, absorption; E, emission.}
    \label{fig:EPR}
\end{figure}

To ensure the alignment of \textbf{\textit{B}}$_\textbf{0}$ with the molecular X-axis of pentacene, the mounting of the pentacene-doped \textit{p}-terphenyl single crystal on the wedged sample holder is carefully designed. Here, we adopt a method such that when the crystal's cleavage plane, i.e. the ab plane, is mounted on the wedged surface of the holder, the molecular plane constructed by the common X-axis and the Y$_\textrm{1}$-axis of a group of the pentacene molecules is horizontal (Fig.~\ref{fig:EPR}a). Thus, if the sample holder, denoted the "XY wedge", is rotated around the lab Z-axis (also the molecular Z$_\textrm{1}$-axis), then we can define an angle $\theta$ that denotes the rotation of the sample holder along its cylindrical axis, and define $\theta=$\SI{0}{\degree} as when \textbf{\textit{B}}$_\textbf{0}$//X. To experimentally achieve it, there are two \textit{macroscopic} angles (shown in Fig.~\ref{fig:EPR}a-c) that are decisive: (i) the angle of the wedge $\alpha$=\SI{15.1}{\degree}; (ii) the angle between the major axis of the wedged elliptical surface and the crystal's b-axis $\beta$=\SI{124}{\degree}. The determination of both angles is aided by the \textit{microscopic} analysis of the relationship between the crystal and molecular coordinate systems\cite{yu1982electron}. As shown in Fig.~\ref{fig:EPR}c, by setting the molecular XY$_\textrm{1}$ plane to be horizontal in the lab frame, $\alpha$ is also the angle between the Z$_\textrm{1}$-axis and the projection of the Y$_\textrm{1}$-axis on the cleavage ab plane (i.e. the wedged surface), say Y$_\textrm{1}$'. Therefore, $\alpha$ is the complementary angle of $\angle$c'Z$_{\textrm{1}}$ which is the angle formed by the crystal c'-axis and the molecular Z$_\textrm{1}$-axis. On the other hand, $\beta$ is the angle between the crystal b-axis and the Y$_\textrm{1}$'-axis that is equal to arctan[cos($\angle$aZ$_\textrm{1}$)/cos($\angle$bZ$_\textrm{1}$)] where the angles are denoted in the same manner as that of $\angle$c'Z$_{\textrm{1}}$. The values of  $\angle$c'Z$_{\textrm{1}}$, $\angle$aZ$_\textrm{1}$ and $\angle$bZ$_\textrm{1}$ can be obtained via the crystallography of the monoclinic \textit{p}-terphenyl lattice\cite{rietveld1970x} ({see Supplementary Section 2}).

We verify the correctness of the sample mounting by implementing the angular-dependent trEPR measurements of the sample. As shown in Fig.~\ref{fig:EPR}d, the overall trEPR spectra measured at different $\theta$ closely reproduce the rotation pattern as well as the polarization [i.e. the absorptive (A)/emissive (E) signals] simulated ({see Supplementary Section 3}) with \textbf{\textit{B}}$_\textbf{0}$ rotated in the pentacene's XY$_\textrm{1}$ molecular plane. The minor deviation observed near \textbf{\textit{B}}$_\textbf{0}$//Y$_\textrm{1}$ might arise from the imperfection in the fabricated wedge angle ({Methods}). Nevertheless, the most critical condition \textbf{\textit{B}}$_\textbf{0}$//X is confirmed by the splitting between the low- and high-field signals at $\theta=$\SI{0}{\degree} or \SI{180}{\degree} matching the theoretical value $(D+3|E|)/(\gamma_\textrm{e}/2\pi)\approx55.6$ mT, where $\gamma_\textrm{e}/2\pi=28$ MHz/mT is the electron gyromagnetic ratio. Furthermore, only one pair of the EPR transitions appears at $\theta=$\SI{0}{\degree} or \SI{180}{\degree} indicating the overlap between the signals of the two different groups of the pentacene molecules that further consolidates the proper alignment of \textbf{\textit{B}}$_\textbf{0}$ with the common X-axis.

\textbf{Characterizations of the solid-state hybrid system}
\begin{figure}[htbp!]
    \centering
    \begin{adjustbox}{width=0.9\textwidth}
        \includegraphics{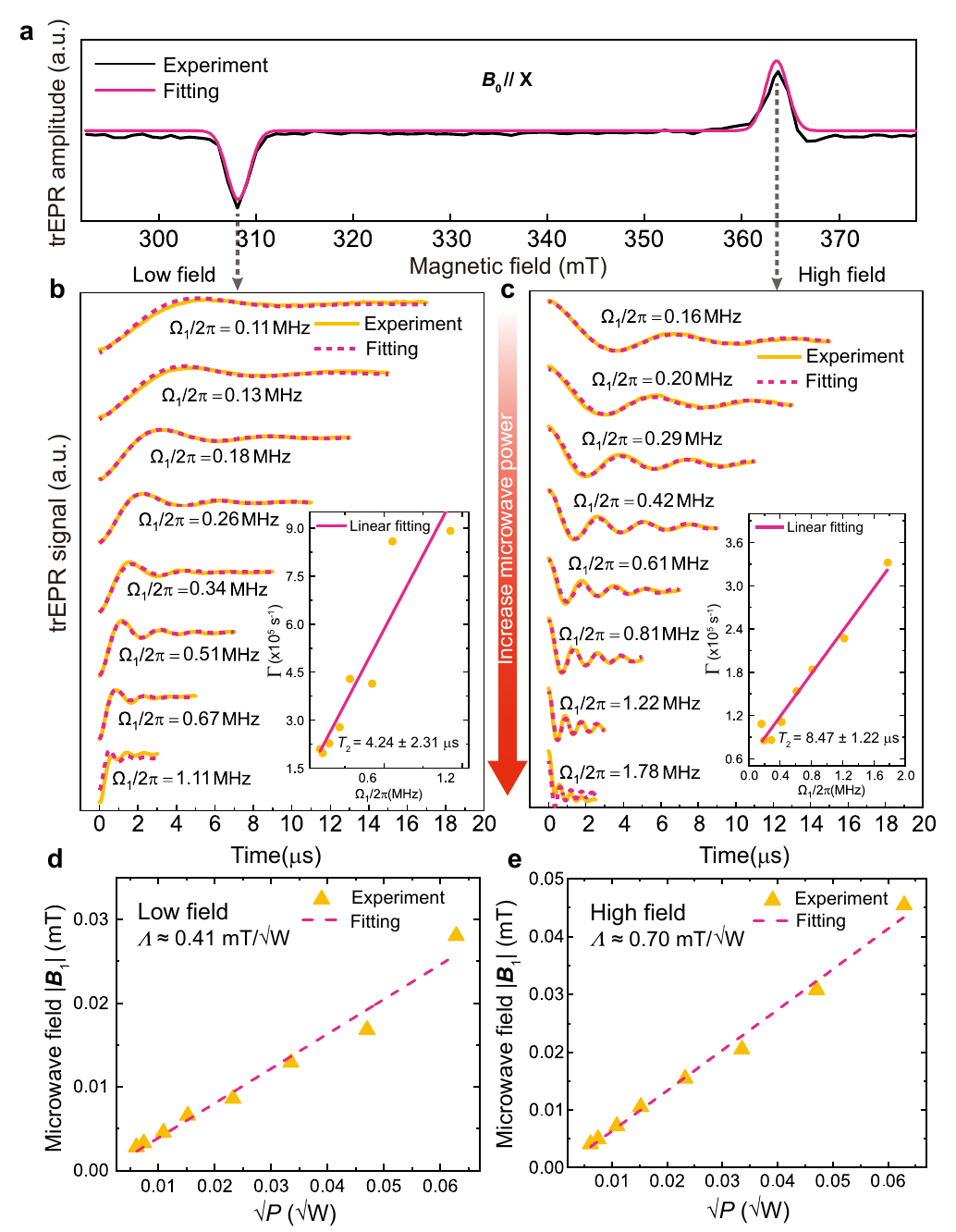}
    \end{adjustbox}
    \caption{\textbf{Characterizations of the pentacene triplet spin dynamics and the resonator at room temperature}. \textbf{a,} trEPR spectrum measured (black) and simulated (pink) under \textbf{\textit{B}}$_\textbf{0}$//X. \textbf{b,c,} Microwave-power-dependent transient spin dynamics of pentacene triplets measured (yellow) at the EPR line positions of \textbf{a}. The microwave power is increased from the top to bottom traces and consistent for both low- and high-field measurements. Pink dashed curves: fittings of the transient spin dynamics. Insets: damping rates of the Rabi oscillations $\Gamma$ as a function of the Rabi frequencies $\Omega_{1}/2\pi$. Linear fittings with the correlation $\Gamma = 1/2T_{2}+\varepsilon\Omega_{1}/2\pi$ provide the decoherence time $T_{2}$ of pentacene triplets at low and high fields. \textbf{d,e,} Determination of the resonator's conversion factors $\mathit{\Lambda}$ at low and high fields. The conversion factors are obtained by linearly fitting the microwave fields |\textbf{\textit{B}}$_\textbf{1}$| (deduced from the Rabi frequencies in \textbf{b} and \textbf{c}) to the square roots of the microwave powers $\sqrt{P}$ inputting the resonator. }
    \label{fig:different power}
\end{figure}

With \textbf{\textit{B}$_\textrm{0}$} and the pentacene-doped \textit{p}-terphenyl crystal properly set up in the device, we characterize the parameter space of the hybrid system \textit{in situ} using the trEPR technique. The maser emission rate $\kappa_\textrm{m}$ is known to be proportional to the stimulated transition probability $W_{0\leftrightarrow-1}$ between the triplet sublevels T$_\textrm{0}$ and T$_\textrm{-1}$ that, for the resonant transition, $W_{0\leftrightarrow-1}\propto(\gamma_\textrm{e}|\textbf{\textit{B}}_\textbf{1}|)^2T_\textrm{2}$\cite{siegman1964microwave} and the squared term is associated with the resonator's conversion factor $\mathit{\Lambda}$. Thus, the system parameter space studied here contains two parameters crucial for the performance of the maser device, which are the spin decoherence time $T_\textrm{2}$ and the resonator's conversion factor $\mathit{\Lambda}$. 

At the low- and high-field EPR transition positions shown in Fig.~\ref{fig:different power}a, we perform the trEPR measurements as a function of the microwave power to investigate the coherence properties of the pentacene triplet spins at room temperature. Upon increasing the microwave power irradiating the sample $P$, the Rabi oscillations start to occur and the Rabi frequency $\Omega_\textrm{1}$ is enhanced (Fig.~\ref{fig:different power}b,c). Meanwhile, we observe the damping of the Rabi oscillations becomes faster with the increased microwave power. According to the previous studies\cite{boscaino1993non,de2012quantum}, the damping rates of the Rabi oscillations $\Gamma$ [which can be obtained by fitting the trEPR signals ({Methods})] have two contributions, $\Gamma = 1/2T_{2}+\varepsilon\Omega_{1}/2\pi$. The former is the intrinsic decoherence rate of the spins, whereas the latter is controlled by the Rabi frequency and $\varepsilon$ is a constant. As shown in the insets of Fig.~\ref{fig:different power}b,c, the linear correlation between $\Gamma$ and $\Omega_\textrm{1}$ is verified by the linear fitting of our results, in which the spin decoherence time $T_\textrm{2}$ and $\varepsilon$ are set to be the fitting parameters. Based on the above analysis, the decoherence time of the pentacene triplet spins is obtained, which reveals a significant difference between the low- and high-field results. The decoherence time in the low field ($4.24 \pm 2.31 \, \mu$s) is only half of that measured in the high field. Additionally, with the same applied microwave power, the Rabi frequencies in the low field are also smaller to those in the high field. The similar phenomena have been reported in the electron spin echo (ESE) experiments with the same spin system and attributed to the distinct conditions of the hyperfine interactions in the low and high fields\cite{kouskov1995pulsed}. In the low field, the nonsecular term of the hyperfine interactions is dominating which promotes the mixing and shuffle among the hyperfine isochromats and thus the resultant disorder of the effective magnetic field leads to degradation of the electron spin coherence and smearing-out of the Rabi oscillations. In contrast, for the high-field transition, the contribution of the nonsecular term is minor so that the precessions of the nuclear and electron spins are adiabatic with respect to the external field giving rise to the longer electron spin coherence as well as the higher effectiveness of the applied microwave field as an electron-spin rotation operator.

The Rabi oscillations observed in the trEPR signals can also provide the information of the resonator's conversion factor $\mathit{\Lambda}$ which quantifies the efficiency of the microwave power-to-field conversion in a resonator and defined by $\mathit{\Lambda}=|\textbf{\textit{B}}_\textbf{1}|/\sqrt{P}$\cite{hyde1989multipurpose}. As for the pentacene triplet spins, the Rabi frequency $\Omega_{1} =\sqrt{2}\gamma_\textrm{e}|\textbf{\textit{B}}_\textbf{1}|$ where the factor $\sqrt{2}$ arises from the spin number $S=1$ of pentacene\cite{weiss2017strongly}, the conversion factor can be rewritten to be $\mathit{\Lambda} = \Omega_\textrm{1}/(\gamma_\textrm{e}\sqrt{2P})$. By plotting the measured $\Omega_\textrm{1}$ against $\sqrt{P}$ together with linear fittings (Fig.~\ref{fig:different power}d,e), we obtain the conversion factors from the fitted slopes which are $\mathit{\Lambda} \approx 0.41 \ \mathrm{and} \ 0.70 \ \mathrm{mT/\sqrt{W}}$
 for the low- and high-field measurements, respectively. In general, the conversion factor, as an intrinsic parameter characterzing the specific mode of a resonator, should not be affected by the condition of the external magnetic field. Our finding implies the spin coherence properties that affect the measured Rabi frequencies can lead to misinterpretation of the resonator's conversion factor and thus need to be clarified in advance. Here, owing to the smaller influence of the hyperfine interactions on the electron spin flip-flop driven by $\textbf{\textit{B}}_\textbf{1}$ in the high field, we take the high-field result $\mathit{\Lambda}=0.70$ mT/$\mathrm{\sqrt{W}}$ as the rational conversion factor of our resonator which outperforms the sapphire resonators ($\mathit{\Lambda}=0.42$ mT/$\mathrm{\sqrt{W}}$\cite{weber2001elexsys}) extensively used in the commercial X-band EPR spectrometers.

\textbf{Microwave quantum amplification}

\begin{figure}[htbp!]
    \centering
    \begin{adjustbox}{width=0.75\textwidth}
        \includegraphics{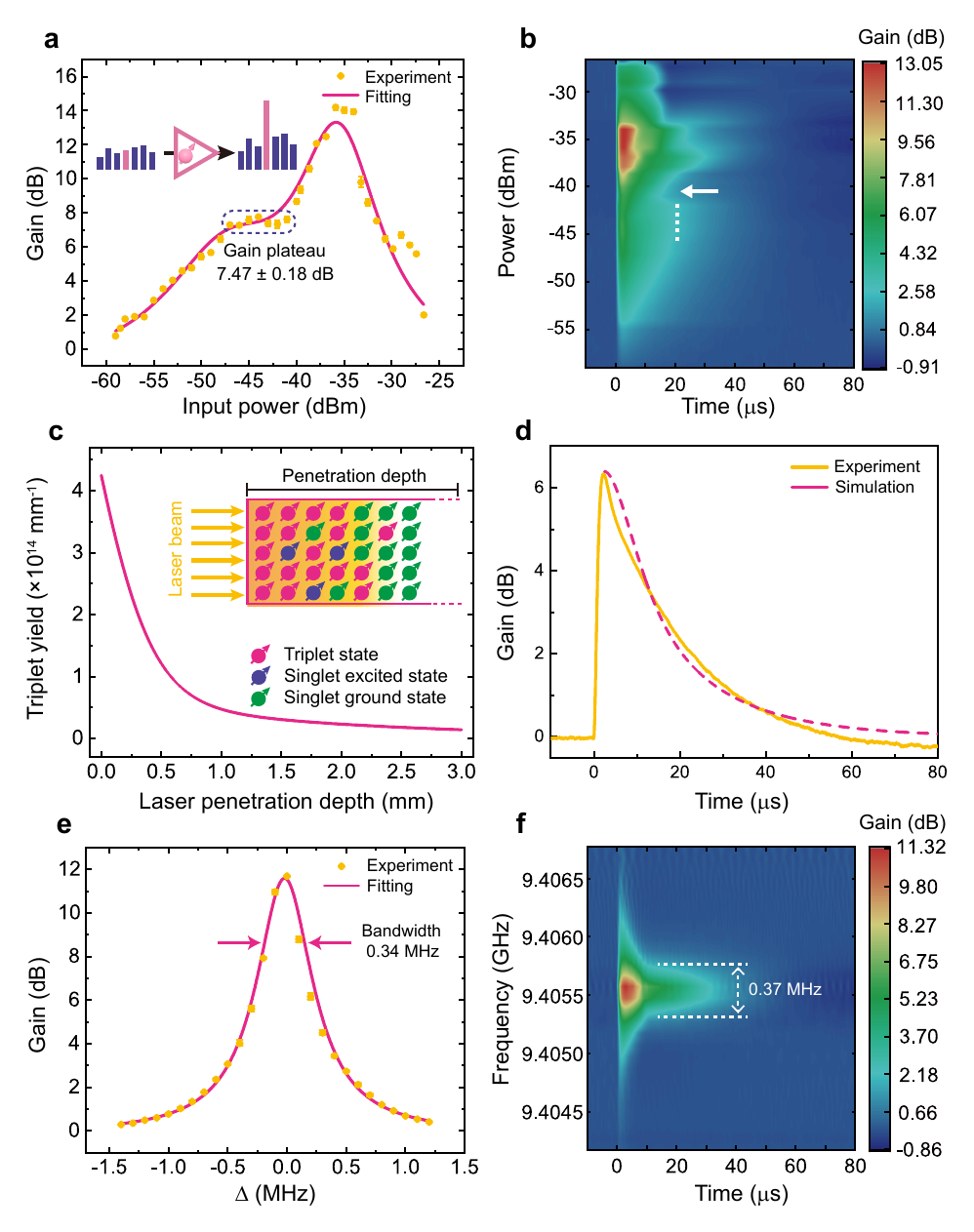}
    \end{adjustbox}
    \caption{\textbf{Performance of the microwave quantum amplifier at room temperature.} \textbf{a,} Dependence of the gain on the power of the input microwave signal at 9.4056 GHz. The means (yellow) of the measured gains at different input powers are fitted (pink) to a double-Lorentzian function. Error bars, standard deviations of the gain values obtained in five individual measurements at a specific input power. The input power region with almost constant gains is highlighted using a dashed box with the averaged gain and its standard deviation labelled. \textbf{b,} Time-resolved quantum amplification of the input microwave signals with varying powers. Dashed line, the power regime with the same amplification durations. Arrow: the turnover point of the amplification durations. \textbf{c,} Simulated sum of the spin populations in the triplet sublevels T$_\textrm{0}$ and T$_\textrm{-1}$ per depth as a function of the depth of a 590-nm pulsed laser penetrating the pentacene-doped \textit{p}-terphenyl crystal with a pump intensity of 23.87 mJ cm$^{-2}$ and duration of 7 ns. Inset: schematic evolution of the pentacene molecular spins at different sample depths when a pulsed laser beam illuminates the crystal. \textbf{d,} Experimental (yellow) and simulation (pink) results of the time-resolved quantum amplification of a 9.4056-GHz input signal with a power of -46 dBm. \textbf{e,} Dependence of the gain on the detuning $\Delta$ of a -35.9-dBm input microwave signal from the central frequency of the resonator. The means (yellow) of the measured gains at different detunings are fitted (pink) to a Lorentzian function with the associated gain bandwidth labelled. Error bars, standard deviations of the gain values obtained in five individual measurements at a specific detuning. \textbf{f,} Time-resolved quantum amplification of the input microwave signals with distinct frequencies. The frequency spacing between the two turnover points of the amplification durations is labelled.}
    \label{fig:maser amplifier}
\end{figure}

After the characterizations of the hybrid system, we first investigate its quantum amplification features under \textbf{\textit{B}}$_\textbf{0}$//X. The device by itself remains the same as in the trEPR measurements including the one-port reflection scheme with the port critically coupled, but we employ a higher pulsed optical pump intensity of 23.87 mJ cm$^{-2}$ and the measurement circuit as shown in {Supplementary Fig.1}. At the position where the strongest low-field emissive trEPR signal emerges (i.e. |\textbf{\textit{B}}$_\textbf{0}$|$\sim$307 mT in Fig.~\ref{fig:EPR}a), we measure the gain of the maser-based quantum amplifier as a function of the microwave input power with a frequency $\omega_\textrm{d}$ resonant with the hybrid system, i.e. $\omega_\textrm{d} = \omega_\textrm{c} = \omega_\textrm{s} =2\pi\times$9.4056 GHz. Fig.~\ref{fig:maser amplifier}a shows the maximum gain of $\sim14$ dB can be achieved with the input power of -35.9 dBm and the higher microwave power results in a clear saturation behaviour. We also notice that the gain stays relatively stable at $7.47\pm0.18$ dB within the input power ranging from -47 to -41 dBm. The gain plateau enhances the practicality of the amplifier due to the suppression of the non-linear output response which is difficult for calibrations. Since the maser device operates in a pulsed mode, we further analyze the dynamical amplification process as shown in Fig.~\ref{fig:maser amplifier}b. Interestingly, the duration of the amplification process also varies with the input power and closely resembles the trend of the gain in Fig.~\ref{fig:maser amplifier}a. The duration increases up to 30 $\mu$s until reaching the point where the gain saturation appears. The amplification lasts nearly the same within the gain plateau and a sudden reduction of the duration arises when the input power is increased to -41 dBm, which is as well one of the turnover points of the gain tendency shown in Fig.~\ref{fig:maser amplifier}a.

Due to the special characteristics of the gain and its dynamics within the gain plateau, we conduct the theoretical analysis of the results measured in the regime. First, we calculate the number of the inverted pentacene triplet spins participating in the quantum amplification process based on the light propagation theory\cite{quan2023general} (see {Supplementary Section 4} for more details). Fig.~\ref{fig:maser amplifier}c reveals that instantaneously after the 7-ns optical pumping, the triplet spins generated in the crystal is not uniformly distributed along the light penetration depth. We obtain the total yield of the spin populations $N_\textrm{total}$ in the mostly considered triplet sublevels T$_\textrm{0}$ and T$_\textrm{-1}$ to be $2.1\times10^{14}$ by the integration of the curve in Fig.~\ref{fig:maser amplifier}c. As the instantaneous polarization of the two-level spin system is 0.73, the inverted spin number $\Delta N=1.5\times10^{14}$ can be obtained. In our system, since the spin resonance linewidth $\Delta \omega_\textrm{s}/2\pi$ measured with the trEPR shown in Fig.~\ref{fig:different power}a is 64.73 MHz which is significantly larger than the resonator bandwidth $\Delta \omega_\textrm{c}/2\pi = 0.85$ MHz, we here define the ratio $R = \Delta\omega_\textrm{c}/\Delta\omega_\textrm{s}$ for calibrating the number of the actual inverted spins $\Delta N' = R\Delta N = 2\times10^{12}$ that can contribute to the masing process. By incorporating the calibrated inverted triplet spin number into a driven Maxwell-Bloch model ({Methods}), we quantitatively reproduce the amplification process as shown in Fig.~\ref{fig:maser amplifier}d which further consolidates the adequacy of the framework of cavity quantum electrodynamics (cavity QED) for describing the coherent microwave emission from the solid-state hybrid quantum systems containing collective
spin ensembles\cite{rose2017coherent,angerer2018superradiant,breeze2018continuous,breeze2017room,zhang2022cavity}. 

In addition to investigating the gain dependence on the input microwave power, we also study the effect of the input frequency on the gain of the quantum amplifier. As shown in Fig.~\ref{fig:maser amplifier}e, by fixing the resonator frequency at the spin resonance of 9.4056 GHz and varying the detuning $\Delta = (\omega_\textrm{d}-\omega_\textrm{s})/2\pi$ of the microwave input with a power of -35.9 dBm, we obtain a Lorentzian distribution of the amplifier gain over the input detunings and the amplifier bandwidth $BW$ can thus be fitted to be 0.34 MHz. The peak gain of 12 dB is slightly reduced compared with that observed in Fig.~\ref{fig:maser amplifier}a which might arise from the resonator frequency drift or the fluctuations of the optical pump intensity. On the other hand, we find the dynamics profile of the amplification process shown in Fig.~\ref{fig:maser amplifier}f is highly symmetrical around the resonance analogous to the gain-detuning correlation demonstrated in Fig.~\ref{fig:maser amplifier}e and the amplification durations abruptly increase from 10 to 40 $\mu$s when the input frequency is tuned approaching the resonance. The spacing between the turnover points of the durations on both sides of the resonance is measured to be 0.37 MHz that is close to the amplifier bandwidth. Indeed, the turnover points observed here together with the one labelled in Fig.~\ref{fig:maser amplifier}b may indicate unexplored room-temperature cavity QED phenomena, whose origins deserve more in-depth theoretical analysis in the future work.

Furthermore, to consolidate our claim of realizing the maser-based quantum amplification, we determine the relationship of the rate of maser emission $\kappa_\textrm{m}$ to the specific loss rates $\kappa_\textrm{0}$ and $\kappa_\textrm{e}$ associated with the resonator. All above rates can be characterized respectively by the reciprocal values of their corresponding quality factors: the magnetic, unloaded and external quality factors $Q_\textrm{m}$, $Q_\textrm{0}$ and $Q_\textrm{e}$, based on $\kappa_\textrm{m,0,e} = \omega_\textrm{m,0,e}/Q_\textrm{m,0,e}$, where $\omega_\textrm{m,0,e}$ represents the associated frequencies that are generally identical to the resonator frequency (or spin transition frequency if on resonance). For our quantum amplifier exploiting the one-port (reflection) maser scheme, the output (which is also the input) is critically coupled that leads to $Q_\textrm{0}=Q_\textrm{e}= 2.2\times10^4$. The magnetic quality factor is defined as following\cite{siegman1964microwave}: 
\begin{equation}\label{Qm}
    Q_\textrm{m}=\frac{1}{\gamma_\textrm{e}^2\mu_\textrm{0}\hbar\Delta n\sigma^2\eta T_\textrm{2}}
\end{equation}
where $\mu_\textrm{0}$ is the permeability of free space, $\Delta n = \Delta N'/V_\textrm{crystal} = 3.3\times10^{20}$ $\textrm{m}^{-3}$ is the actual inverted triplet spin \textit{density} contributing to the masing process ($V_\textrm{crystal}=6$ mm$^3$ is the volume of the pentacene-doped \textit{p}-terphenyl crystal), $\sigma^2 = 0.5$ is the normalized transition probability matrix element for the spin $S=1$ system when the $\textbf{\textit{B}}_\textbf{1}$ field is linearly polarized, $\eta = V_\textrm{crystal}/V_\textrm{mode} = 0.027$ is the filling factor and $T_\textrm{2}$ = $4.24 \pm 2.31 \, \mu$s is the spin decoherence time measured with the trEPR. Therefore, we can obtain the value of $Q_\textrm{m} \approx 1.3\times10^4$ which fulfills the criteria for maser amplification that $\kappa_\textrm{0}<\kappa_\textrm{m}<\kappa_\textrm{0}+\kappa_\textrm{e}$, i.e. $Q_\textrm{0}^{-1}<Q_\textrm{m}^{-1}<Q_\textrm{0}^{-1}+Q_\textrm{e}^{-1}$\cite{breeze2015enhanced}. Moreover, by knowing $Q_\textrm{m}$, we can obtain the theoretical values of the amplifier gain $G_\textrm{cal}^\textrm{dB}$ and bandwidth $BW_\textrm{cal}$ according to\cite{siegman1964microwave}:
\begin{equation}\label{gaincal}
    G_\textrm{cal}^\textrm{dB} = 10\lg(\frac{Q_\textrm{e}^{-1}-Q_\textrm{0}^{-1}+Q_\textrm{m}^{-1}}{Q_\textrm{e}^{-1}+Q_\textrm{0}^{-1}-Q_\textrm{m}^{-1}})^2
\end{equation}
\begin{equation}\label{bwcal}
    BW_\textrm{cal}=\omega_\textrm{0} (Q_\textrm{0}^{-1}+Q_\textrm{e}^{-1}-Q_\textrm{m}^{-1})/2\pi
\end{equation}
which are $G_\textrm{cal}^\textrm{dB} = 14.8$ dB and $BW_\textrm{cal} = 0.13$ MHz, respectively. $G_\textrm{cal}^\textrm{dB}$ shows a good agreement with the maximum gain $\sim 14$ dB obtained in our experiments, whereas the experimental amplifier bandwidth $BW=0.34$ MHz is almost three times wider than the theoretical result which could be attributed to the effect of the spin inhomogeneous broadening. The noise temperature of amplifiers is also a highly critical parameter for evaluating the amplifier performance. Due to the pulsed operation of our amplifier, the routine Y-factor measurements with the commercially available noise analyzer can not applied here. Thus, we evaluate the optimal noise temperature of our amplifier $T_\textrm{a}$ based on the noise theory in which the circulator and transmission lines connected with the device are assumed to be ideal (i.e. lossless and perfectly matched)\cite{siegman1964microwave}:
\begin{equation}\label{Tcal}
    T_\textrm{a} \approx |T_\textrm{s}| + (Q_\textrm{m}/Q_\textrm{0})T_\textrm{bath}
\end{equation}
where $T_\textrm{s} \sim -0.24$ K is the spin temperature derived from the definition $\tanh{(\hbar\omega_\textrm{s}/2k_\textrm{B}T_\textrm{s})}=\frac{P_\textrm{-1}-P_\textrm{0}}{P_\textrm{-1}+P_\textrm{0}}$\cite{siegman1964microwave} in which $k_\textrm{B}$ is the Boltzmann constant and $T_\textrm{bath}=290$ K is the thermal bath (ambient) temperature. Combined with the values of $Q_\textrm{m}$ and $Q_\textrm{0}$, we can estimate the noise temperature of our amplifier $T_\textrm{a}\approx172$ K corresponding to a noise figure of about 2.02 dB. It is apparent that at room temperature, the noise temperature of the maser-based quantum amplifier is dominated by the second term of Eq.~(\ref{Tcal}), thus, the noise temperature can be further reduced by (i) adjusting the system parameters in Eq.~(\ref{Qm}) to minimize $Q_\textrm{m}$ and (ii) employing the emerging spin refrigerators\cite{wu2021bench,ng2021quasi,zhang2022microwave,fahey2023steady,blank2023anti} to cool the thermal bath. If the lowest $T_\textrm{bath} = 50$ K is inserted into Eq.~(\ref{Tcal}) which was experimentally achieved at room temperature using the pentacene-based spin refrigerator\cite{wu2021bench}, the noise temperature can be improved to $T_\textrm{a}\approx30$ K corresponding to a substantially reduced noise figure of 0.43 dB which will surpass the conventional room-temperature X-band low-noise amplifiers (LNAs) based on parametric amplifiers (paramps), field-effect transistors (FETs) and high electron-mobility-transistors (HEMTs)\cite{whelehan2002low}.

\textbf{Microwave quantum oscillation} 
\begin{figure}[htbp!]
    \centering
    \begin{adjustbox}{width=\textwidth}
        \includegraphics{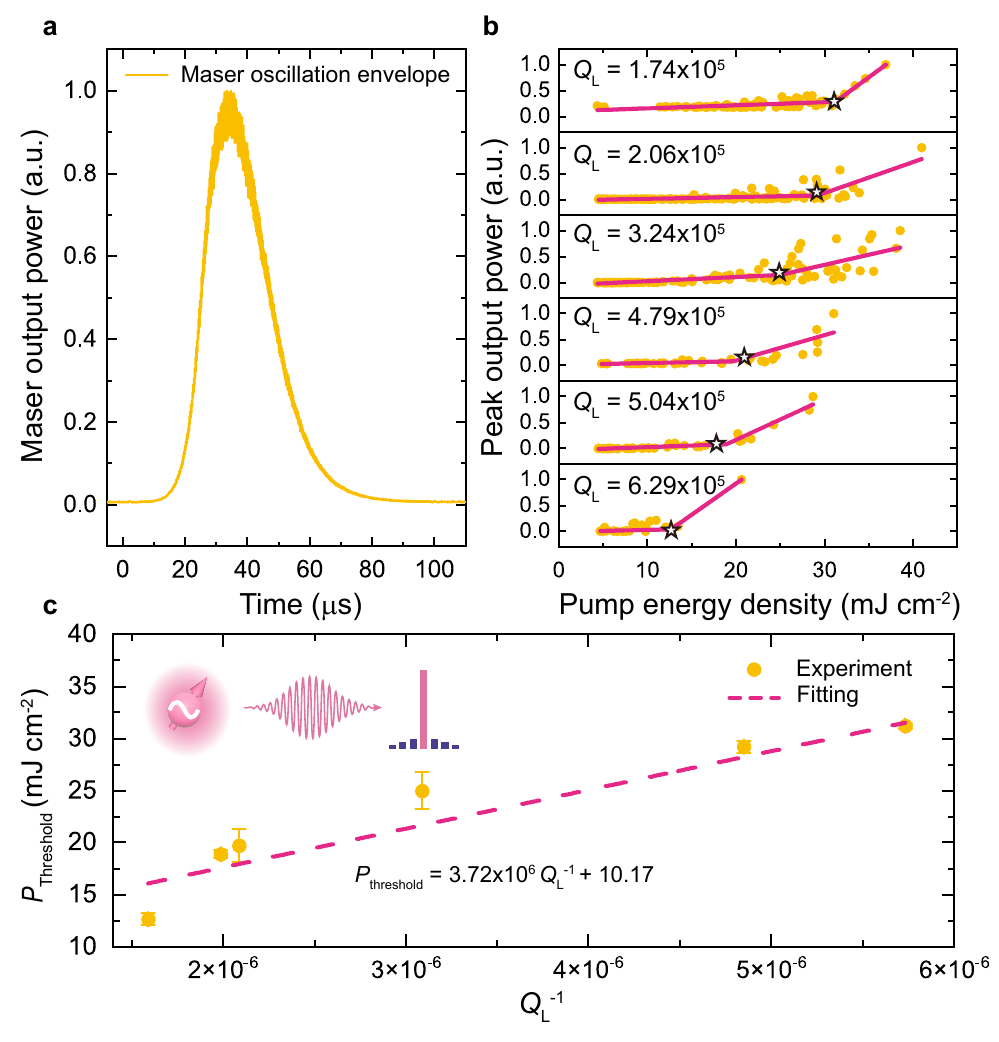}
    \end{adjustbox}
    \caption{\textbf{Performance of the microwave quantum oscillator at room temperature.} \textbf{a,} Instantaneous output power of the 9.4043139-GHz quantum oscillator with a boosted quality factor of $5.0\times10^{5}$ and a pump intensity of 16.54 mJ cm$^{-2}$. \textbf{b,} Threshold measurements of the quantum oscillator with different quality factors. Piecewise linear fittings (pink lines) are used to extracted the thresholds (star). \textbf{c,} Linear dependence of the maser threshold on the reciprocal loaded quality factor.}
    \label{fig:maser oscillator}
\end{figure}

To achieve the quantum oscillation, it is required to meet the criteria that $\kappa_\textrm{m}>\kappa_\textrm{0}+\kappa_\textrm{e}$, i.e. $Q_\textrm{m}^{-1}>Q_\textrm{0}^{-1}+Q_\textrm{e}^{-1}=Q_\textrm{L}^{-1}$. To this end, compared to the experimental conditions of the quantum amplification discussed above, we can either decrease $Q_\textrm{m}$ by using stronger optical pumping for increasing the inverted triplet spin density $\Delta n$ or lower the loaded quality factor $Q_\textrm{L}$ of the resonator. In our attempts, we could not realize the quantum oscillation with the former strategy even with the maximum accessible pump intensity. Hence, we choose to modify $Q_\textrm{L}$ by connecting the maser device with a feedback loop in which the power dissipation in the resonator can be controlled on demand by adjusting the gain and phase in the circuitry (see {Supplementary Fig.1}), resulting in a rather wide tunable range of $Q_\textrm{L}$ from the native $1.1\times10^4$ to about $6.5\times10^5$, above which the undesired classical self-oscillation of the circuitry instead of the quantum oscillation will occur. Such a $Q$-boosting approach has benefited the recent room-temperature cavity QED studies on a variety of quantum systems, such as the solid-state electron spins\cite{wu2022enhanced,ng2021exploring,attwood2023n}, magnons\cite{yao2017cooperative,yao2023coherent} and vapor atoms\cite{jiang2021floquet}.

With a pump intensity of 16.54 mJ cm$^{-2}$ and a boosted $Q_\textrm{L}\sim5.0\times10^{5}$, Fig.~\ref{fig:maser oscillator}a shows a typical quantum oscillation output (power envelope) from the maser device with an on-resonance frequency of 9.4043139 GHz. The output peak latency arises from the build-up time for the intra-resonator microwave photons to surpass the masing threshold as well as the rise time of the resonator $\tau_\textrm{R}=2Q_\textrm{L}/\omega_\textrm{c}\approx17~\mu$s. We systematically investigate the effect of $Q_\textrm{L}$ on the quantum oscillation by measuring the maser thresholds at different $Q_\textrm{L}$ values as shown in Fig.~\ref{fig:maser oscillator}b,c. The results show the reduction of the maser threshold as $Q_\textrm{L}$ is increased and the inverse proportionality revealed by the fitting agrees well with the theoretical predictions\cite{breeze2015enhanced,breeze2018continuous,zollitsch2023maser}. According to the fitted function, with the native $Q_\textrm{L}=1.1\times10^4$, the optical pump intensity required for achieving the maser threshold $P_\textrm{threshold}$ could be $\sim348$ mJ cm$^{-2}$ which is significantly beyond the capacity of our pump source. Therefore, the $Q$-boosting approach offers the advantages of (i) reducing the optical pumping requirement for the device and (ii) providing reliable prediction of the maser threshold with a native $Q_\textrm{L}$, which is an important step towards the practical usage of the maser-based quantum oscillator and can also facilitate the exploration and evaluation of novel maser gain media owing to the elimination of the requirement for fabricating high-$Q$ resonators in early-stage development.

\section*{Discussion}
\hspace{2em} In this work, by coupling optically polarized pentacene triplet spins to a sapphire resonator, we have demonstrated a solid-state hybrid quantum system that functions as a quantum electronic device for masing, which is a form of coherent microwave emission, in the X-band at room temperature. Our study unveils the suitability and convenience of the trEPR technique for \textit{in situ} characterizations of the active components of the device, by which we observe a relatively long $T_\textrm{2}$ of the pentacene triplet spins, up to $\sim$8.5 $\mu$s at room temperature, showing their potential to be applied in the nascent field of the (organic) molecular spin-based QIPC\cite{wasielewski2020exploiting,gaita2019molecular}. We emphasize that compared with the pulse EPR, the trEPR, as a continuous-wave EPR technique, provides an alternative and more direct approach for evaluating the decoherence of the optically induced paramagnetic spins. In terms of the device functionalities, we have shown the maser emission can be exploited for quantum amplification and oscillation, of which the performance can be tuned on demand via the external microwave driving and the active power-dissipation control. For the future improvements, we envision that the bench-top spin refrigeration technique\cite{wu2021bench,ng2021quasi,fahey2023steady,zhang2022microwave} can be employed for reducing the noise temperature of the device as a quantum amplifier. Additionally, future work can be conducted to lower the threshold of the quantum oscillation by minimizing the ohmic loss of the resonator's copper cavity via silver and gold coating\cite{zollitsch2023maser}. From the material perspective, the longer $T_\textrm{2}$ and the higher inverted spin density can reduce the masing threshold, for which the deuteration of pentacene\cite{kouskov1995pulsed} and the invasive optical pumping\cite{wu2020invasive} can be implemented, respectively. This experimental work also reveals intriguing dynamics of the hybrid quantum system, e.g. the phase-transition-like behavior observed during the transient quantum amplification, which may spur the theoretical investigations on the room-temperature driven cavity QED of such a hybrid quantum system involving an inhomogeneously broadened spin ensemble in the microwave regime.

\section*{Methods}
\textbf{Spin Hamiltonian} \\
The spin Hamiltonian\cite{hintze2017photoexcited} used for the pentacene triplet electrons is 
\[{H}=D({{{S}}_{\mathrm{z}}}^{2}-\frac{1}{3}\textbf{\textit{S}}^{2})+E({{{S}}_{\mathrm{x}}}^{2}-{{{S}}_{\mathrm{y}}}^{2})+g_\textrm{e}\mu_\textrm{B} \textbf{\textit{B}}_\textbf{0}\cdot\textbf{\textit{S}}\]
where the parameters $D$ and $E$ characterize the zero-field splitting, $g_\textrm{e}\approx2.0$ is the electron $g$-factor, $\mu_\textrm{B}$ is the Bohr magneton and \textbf{\textit{S}}\ is the triplet spin eigenvector. The first two terms in the Hamiltonian describe the zero-field splitting, representing the dipolar interaction between two unpaired electron spins. The third term considers the Zeeman interaction induced by an external magnetic field, causing additional energy-level splittings. Here, we neglect all contributions from the nucleus in the Hamiltonian. When the external magnetic field is aligned with the molecular X-axis, the eigenvalues of the spin Hamiltonian, i.e. the energy levels, are given by the following equation\cite{hintze2017photoexcited}:
\begin{equation}
\textbf{\textit{B}}_\textbf{0}//\textrm{X}:{{E}_{\pm 1}}=-\frac{1}{2}\left(\frac{D}{3}-E\right)\pm \sqrt{{\left(\frac{D+E}{4}\right)^2}+(g_\textrm{e}\mu_\textrm{B}|\textbf{\textit{B}}_{\textbf{0}}|)^2}, \quad {E}_\textrm{0}=\frac{1}{3}D-E \notag
\end{equation}
\noindent Therefore, the transition frequency between the triplet sublevels T$_\textrm{0}$ and T$_\textrm{-1}$ corresponds to \(E_\textrm{0}-E_\textrm{-1}\).

\noindent \textbf{Sample preparation and mounting} \\
The sample used in the experiment was a pentacene-doped \textit{p}-terphenyl crystal with a doping concentration of 1000 p.p.m. The crystal-growth process can be referred to the previous study\cite{wu2022enhanced}. First, the as-grown single crystal was cut to obtain the cleavage plane (i.e. the ab plane) and subsequently polished to achieve a sample size of $2 \ \mathrm{mm} \times 1 \ \mathrm{mm} \times 3 \ \mathrm{mm}$. The a-axis of the crystal was determined by the birefrigence method and the b-axis is thus approximately orthogonal to it. By using the Mercury software, the angles between the specific crystal and molecular axes were measured for determining the values of $\alpha$ and $\beta$ (details are available in the {Supplementary Section 2}). Based on the calculated $\alpha$, a quartz rod with a diameter of 4 mm was polished to achieve the wedged sample holder whose actual wedge angle was determined by an optical microscope ({DYJ-630C}) to be $16^\circ$ which is slightly larger than the desired $\alpha$ due to the fabrication error. The sample was fixed on the holder by the silicone grease ({HOTOLUBE}) and rotated on the wedged surface to align its b-axis according to the calculated $\beta$. Note that, there might be an ambiguity of an angle of $\pi$ for the b-axis alignment, which can be confirmed by the trEPR measurements.

\noindent \textbf{Design of microwave dielectric resonators} \\
The microwave dielectric resonator\cite{breeze2018continuous} comprised a cylindrical oxygen-free copper cavity (inner diameter, 36 mm; inner height, 34 mm; {ZhongNuo Advanced Material (Beijing) Technology}) and a polished sapphire ring (relative permittivity $\varepsilon_\rho = 9.394$; outer diameter, 10 mm; inner diameter, 5.1 mm; height, 6.0 mm;  J-Crystal Photoelectric
Technology, China). The support for the sapphire ring was made of Rexolite. The frequency can be tuned by adjusting the copper tuning screw located at the top of the copper cavity. Two loop antennas were inserted inside to couple microwave signals in and out. For the trEPR and the quantum amplification experiments, only one antenna was used. A 4-mm-diameter hole was drilled on the center of the cavity wall to direct the laser beam onto the crystal. The resonator supports a \texttt{$\mathrm{TE}_{01\delta}$} mode whose unloaded $Q_\textrm{0}$ = $2.2 \times 10^4$ was measured by a microwave analyzer ({Keysight N9917A}) when the resonance was tuned at 9.4056 GHz. Using the COMSOL software, the magnetic mode volume $V_\textrm{mode} = 0.22 \, \text{cm}^3$ was calculated based on the ratio of the magnetic energy stored in the mode to the maximum magnetic field energy density:
\[V_{\text{mode}} = \int\limits_{V} \frac{\left|\textbf{\textit{H}}(\textbf{\textit{r}}) \right|^2}{\left| \textbf{\textit{H}}_{\max }(\textbf{\textit{r}}) \right|^2} \, dV\]\\
\textbf{Optical pumping}\\
The optical pump source throughout the experiments was an optical parametric oscillator (OPO) (Deyang Tech. Inc. BB-OPO-Vis, pulse duration $\sim$7 ns). The wavelength and beam size of the output were 590 nm and 4 mm, respectively. For the trEPR and quantum amplification experiments, the pump energies were set to 0.8 and 3.0 mJ, respectively. {To measure the maser thresholds in the quantum oscillation experiments, the pump energies were varied from 0.3 to 5.1 mJ.}

\noindent \textbf{Tr-EPR spectroscopy} \\
The X-band trEPR spectrometer was home-built whose block diagram can be found in {Supplementary Fig. 1}. The angular-dependent trEPR spectra were collected at different $\theta$ by rotating the home-made goniometer in steps of 10 ± 1°. The data at different magnetic fields were the averaged results of 100 measurements. The rotation pattern was simulated using the pepper function in the EasySpin toolbox\cite{stoll2006easyspin}. The parameters used for the simulation can be found in {Supplementary Section 3}. 

\noindent \textbf{Driven Maxwell-Bloch model} \\
The driven Maxwell-Bloch model used to describe the amplification process is:
\begin{align*}
{H} = \omega_{\mathrm{s}} {S}_{\mathrm{z}} + \omega_{\mathrm{c}} {a}^{\dagger} {a} + g ({a}^{\dagger} {S}_{\mathrm{-}} + {a} {S}_{\mathrm{+}}) + V ({a}^{\dagger} + {a})
\end{align*}
Based on this theoretical framework, the equation of motion for the hybrid system can be derived\cite{rose2017coherent}:
\begin{align*}
 & \langle \dot{{a}}(t)\rangle = -iV - \kappa_{\mathrm{c}} \langle{a(t)} \rangle - i g\langle {S}_{\mathrm{-}}(t) \rangle, \\
  & \langle \dot{{S}}_{\mathrm{-}}(t) \rangle = -\kappa_{\mathrm{s}} \langle {S}_{\mathrm{-}}(t) \rangle + 2i g\langle {a}(t) \rangle \langle {S}_{\mathrm{z}}(t) \rangle, \\
  & \langle \dot{{S}}_{\mathrm{z}}(t) \rangle = i g(\langle {a}^{\dagger}(t) \rangle \langle {S}_{\mathrm{-}}(t) \rangle - \langle {a}(t) \rangle \langle {S}_{\mathrm{+}}(t) \rangle) - \gamma \langle {S}_{\mathrm{z}}(t) \rangle.
\end{align*}
To simplify these equations, all spins were assumed to be identical and the semiclassical limit was adopted by neglecting the correlation between the spin ensemble and the cavity photons: $\langle {a} {S}_{\mathrm{i}} \rangle = \langle {a} \rangle \langle {S}_{\mathrm{i}} \rangle, {\mathrm{i}}=+,-,z $.
$\langle \vphantom{{S}_{\mathrm{-}}} {a}(t) \rangle, \langle {S}_{\mathrm{-}}(t) \rangle \text{ and } \langle {S}_{\mathrm{z}}(t) \rangle$ in the Maxwell-Bloch equation are the expected values, and \(g\) is the single spin-photon coupling strength, with its value being 0.69 Hz, $\kappa_{\mathrm{c}}=\omega_\textrm{c}/Q_\textrm{L}$ represents the total cavity loss rate, $\kappa_{\mathrm{s}} = 2/T_\textrm{2}$ is the spin decoherence rate, \(\gamma=4.5\times10^4\) s$^{-1}$\cite{kawahara2015kinetic} indicates the spin depolarization rate taking both the spin-lattice relaxation and the depopulation from the triplet state back to the singlet state into account. \(a^{\dagger}(a)\) are the creation (annihilation) operators for cavity photons and \(S_{\mathrm{z}}\) represents the collective inversion operator with the initial value of $\Delta N'$. The driving strength is defined as $V$\cite{angerer2019non} :
$ V = \sqrt{{P_\textrm{in} \kappa_{\mathrm{c}}}/{\hbar \omega_{\mathrm{c}}}}$, where
$P_\textrm{in}$ represents the microwave input power, i.e. -46 dBm. The measured output of the amplifier $P_\textrm{out}$ correlates with the cavity photon number $n = \langle a^{\dagger}a\rangle$ based on\cite{breeze2017room}:$n =\ P_\textrm{out}(1 + k)/{\hbar \omega_{\mathrm{c}}\kappa_{\mathrm{c}} k}$, where \textit{k} is the coupling coefficient ($k$ = 1 for the critical coupling).

\noindent\textbf{Acknowledgements} We thank Wern Ng for insightful discussions and comments on the manuscript. H.W. acknowledges financial support from the National Natural Science Foundation of China (12204040) and the China Postdoctoral Science Foundation (YJ20210035, 2021M700439 and 2023T160049). B.Z. acknowledges financial support from the National Natural Science Foundation of China (12374462 and 12004037). M.O. acknowledges financial support from the EPSRC New Horizons (EP/V048430/1). Q.Z. acknowledges financial support from the Innovation Program for Quantum Science and Technology (2023AAA040264).

\noindent\textbf{Author Contributions} K.P.W. and H.W. built the experiment, performed the simulation, analyzed the data and wrote the initial manuscript. K.P.W. performed the measurements and processed the data. H.W. conceived the idea. B.Z., X.R.Y., and J.K.Z. contributed to the data acquisition and analysis. M.O. provided the sample, contributed to the maser experiments, and improved the manuscript. H.W. and Q.Z. supervised the project. All authors contributed to revising the manuscript.

\noindent\textbf{Data Availability} All key data that support the findings of this study are included in the main article and its supplementary information. Additional data sets and raw measurements are available from the corresponding author upon reasonable request.

\noindent\textbf{Competing interests} The authors declare no competing interests.

\printbibliography

\end{document}


\captionsetup[figure]{labelfont={bf},labelformat={default},labelsep=period,name={Supplementary Fig.}}
\captionsetup[figure]{font={stretch=1.2}}

\captionsetup{labelfont={bf}, labelsep=period}
\captionsetup[table]{name={Supplementary Table.}}
\date{} 
\maketitle
\newpage 


\begin{center}
\textbf{Supplementary Section 1 - Experimental measurement circuits}
\end{center}
\begin{figure}[htbp!]
    \centering
    \includegraphics[width=\textwidth]{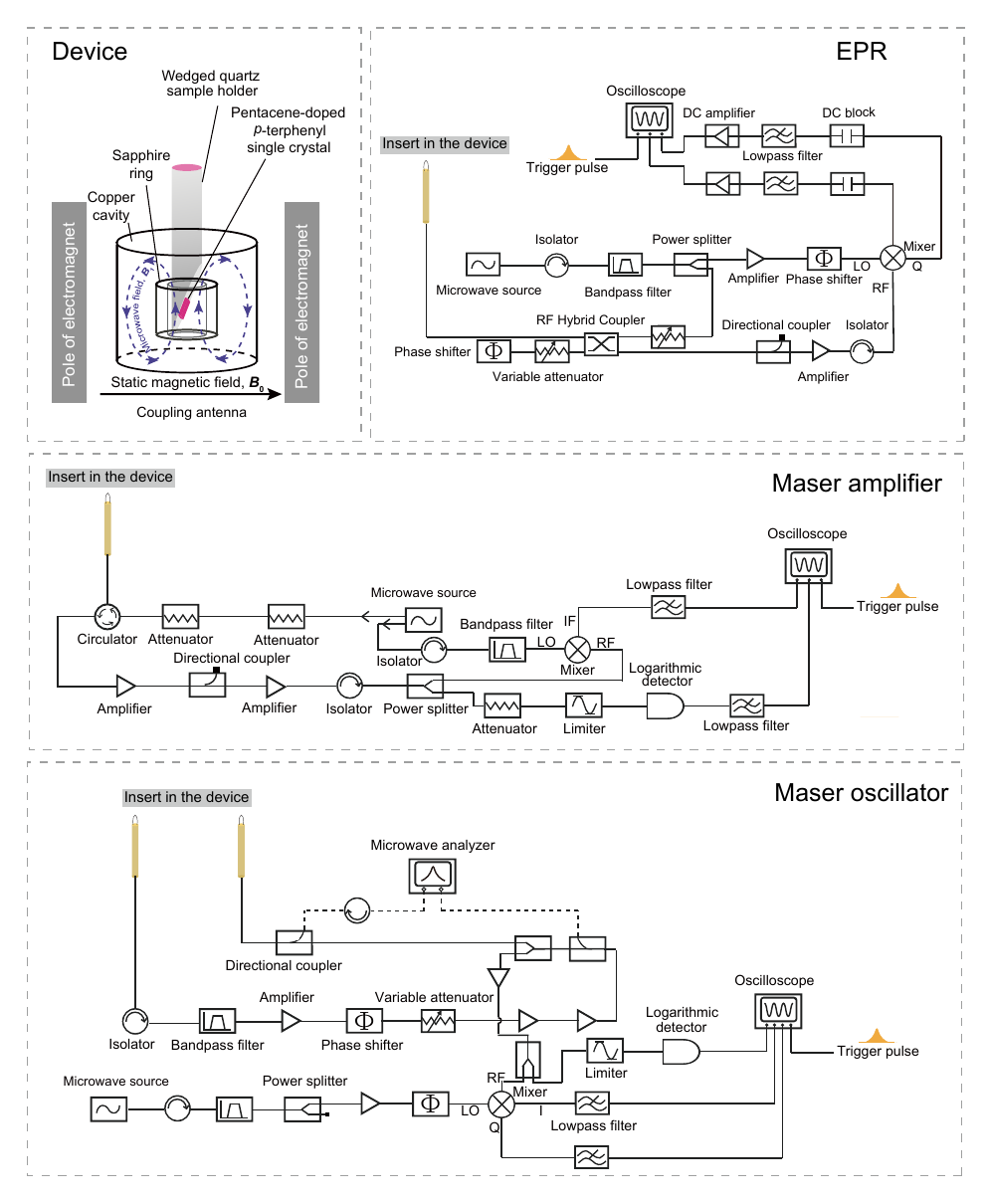}
    \caption{\textbf{Experimental setups.}}
\end{figure}
\FloatBarrier 
\newpage

\begin{figure}[htbp!]
    \centering
    \includegraphics[width=0.9\textwidth]{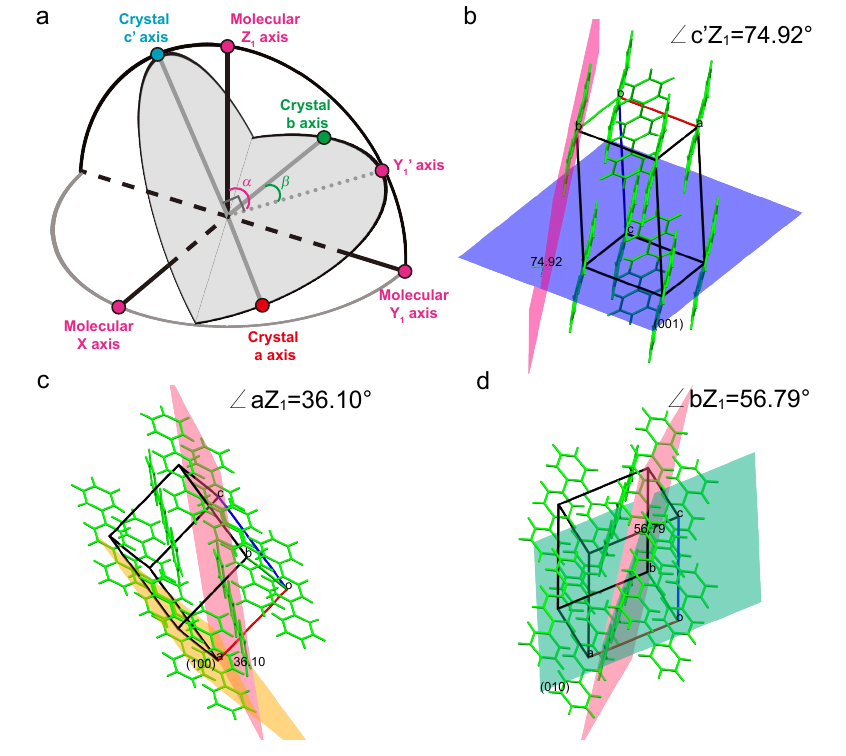}
    \caption{\textbf{Angle measurements.} \textbf{a,} Angular relationship between the crystallographic and molecular axes  \textbf{b-d,} Measurements of the angles \angle c'Z$_1$, \angle aZ$_1$ and \angle bZ$_1$ using the Mercury software.}
    \label{Sfig:angle}
\end{figure}

\begin{center}
\textbf{Supplementary Section 2 - Determination of angles $\alpha$ and $\beta$}
\end{center}
As mentioned in the main text, the values of $\alpha$ and $\beta$ can be calculated according to 
\begin{align}
  & \alpha =90{}^\circ -\mathrm{\angle c \text{'}Z_{1}} \\ 
 & \beta =\arctan \mathrm{\left(\frac{\cos (\angle aZ_{1})}{\cos (\angle bZ_{1})}\right)}  
\end{align}
Therefore, we need to obtain the values of \angle c'Z$_1$, \angle aZ$_1$ and \angle bZ$_1$. To help visualize the above angles, we reuse the Fig.2c in the main text here as Supplementary Fig.~\ref{Sfig:angle}a. To this end, we employ the Mercury software and  the reported crystallography data of \textit{p}-terphenyl\cite{rice2013temperature} to measure the angles between the associated planes to which the above axes are perpendicular. The results are shown in Supplementary Fig.~\ref{Sfig:angle}b-d. By inserting the obtained angles in the above equations, the values of $\alpha$ and $\beta$ can thus be determined to be $15.1{}^\circ$ and $124{}^\circ$, respectively.
\newpage

\begin{center}
\textbf{Supplementary Section 3 - Simulation of the rotation pattern}
\end{center}
The rotation pattern of the EPR spectra shown in the main text was simulated using the pepper function in EasySpin\cite{stoll2006easyspin}.
EasySpin defines a set of standard Cartesian frames. During the simulation, the laboratory, molecular and crystal frames were employed to understand the orientation of the crystal lattice in the spectrometer and the orientation of the pentacene molecules in the \textit{p}-terphenyl crystal. According to the default settings of EasySpin, the laboratory frame's three unit vectors were expressed as $(\mathrm{X_{L}}, \mathrm{Y_{L}}, \mathrm{Z_{L}})$, in which $\mathrm{Z_{L}}$ is horizontal, along the static magnetic field, and $\mathrm{X_{L}}$ is aligned with the microwave field in the resonator, thus vertical. Since the XY$_1$ molecular plane of pentacene-doped \textit{p}-terphenyl is parallel to the static magnetic field, and X, Y$_1$, and Z$_1$ in the molecular frame are perpendicular to each other, we could obtain that the Euler angle from the molecular frame to the laboratory frame is [-90 270 0]. Subsequently, the rotation matrix of the corresponding transformation could be obtained through the Euler angle. Supplementary Table.1 shows the angles and cosines between the molecular axes (X, Y$_1$, Z$_1$) and the crystallographic axes (a, b, c') of \textit{p}-terphenyl, measured using the Mercury software. 

\begin{table}[htbp!]
    \centering
    \begin{tabular}{cccccccccc}
        \hline
        & \angle aX & \angle bX & \angle c'X & \angle aY$_1$ & \angle bY$_1$ & \angle c'Y$_1$ & \angle aZ$_1$ & \angle bZ$_1$ & \angle c'Z$_1$ \\
        \hline
        Angle& 73.08 & 89.62 & 164.75 & 59.04 & 32.67 & 81.26 & 36.10 & 123.21 & 74.92 \\
        \hline
        Cosine of the angle& 0.29 & 0.01 & -0.96 & 0.51 & 0.84 & 0.15 & 0.81 & -0.55 & 0.26 \\
        \hline
    \end{tabular}
    \caption{Angles and cosines}
    \label{tab:angles_cosines}
\end{table}

\noindent The rotation matrix $R$ could thus be obtained through the cosine values. Note that, there might be an ambiguity of an angle of \(\pi\) for the actual sample mounting, which could be confirmed by the trEPR measurements.
\[
R = 
\begin{bmatrix}
    \cos(\mathrm{\angle aX}) & \cos(\mathrm{\angle bX}) & \cos(\mathrm{\angle c\text{'}X}) \\
    \cos(\mathrm{\angle aY_1}) & \cos(\mathrm{\angle bY_1}) & \cos(\mathrm{\angle c\text{'}Y_1}) \\
    \cos(\mathrm{\angle aZ_1}) & \cos(\mathrm{\angle bZ_1}) & \cos(\mathrm{\angle c\text{'}Z_1}) \\
\end{bmatrix}
\]
\noindent According to eulang($R$) $\times$ 180/ \(\pi\) in Matlab, we obtain the Euler angle of the rotation from the crystal frame (a, b, c') to the molecular frame (X, Y$_1$, Z$_1$) which is [326.12 75.70 9.28] $\times$ \(\pi\)/ 180. Multiplication of the two derived rotation matrices could provide the rotation matrix for the transformation from the crystal frame to the laboratory frame, and then the Euler angle of the transformation could be obtained. On the other hand, for the simulation parameters associated with the pentacene triplet spins, the zero-field splitting parameters $D=1395.57$ MHz and $E=-53.35$ MHz\cite{yang2000zero} were employed. The spectrum rotating around the laboratory $\mathrm{X_{L}}$ could be simulated by Easyspin. Due to the uncertainties of the angles measured using the Mercury software and the sample mounting, we found the fine adjustment of the Euler angle of the rotation from the crystal frame (a, b, c') to the molecular frame (X, Y$_1$, Z$_1$) to [326.27 76.445 10.07] × \(\pi\)/180 could offer the better match of the simulated rotation pattern to the experimental results. 

\begin{center}
\textbf{Supplementary Section 4 - The light propagation theory}
\end{center}
\begin{figure}[htbp!]
    \centering
    \includegraphics[width=0.8\textwidth]{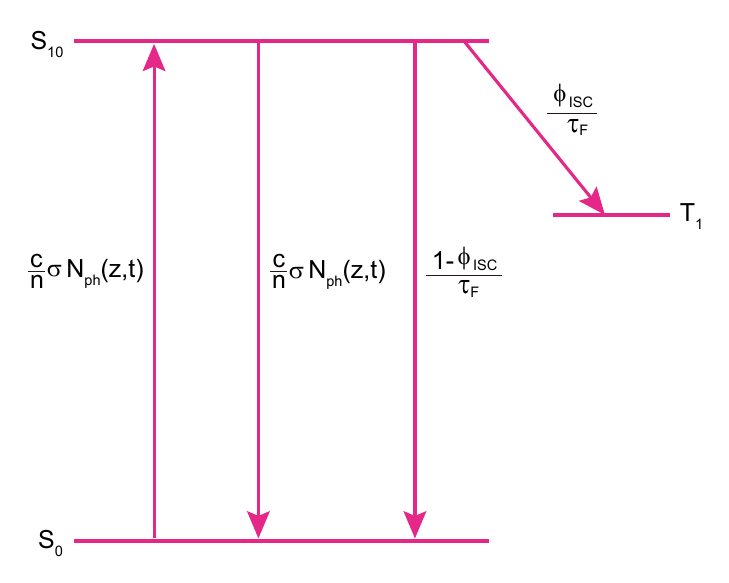}
    \caption{\textbf{Simplified Jablonski diagram and the associated dynamics upon photoexcitation of pentacene molecules.}
    The system is simplified to be a three-level system which involves the singlet ground state S$_{0}$, the singlet excited state S$_\textrm{10}$ and the triplet state T$_\textrm{1}$. The dynamics processes among the electronic states are the optical pumping, the stimulated emission, the spontaneous emission and the intersystem crossing. The rates of the former two processes can be assumed to be the same\cite{carmichael1999statistical} and associated with the time- and penetration-depth-dependent optical photon densities as well as the refractive index and the absorption cross section of the crystal. The rates of the later two processes depend on the intersystem crossing yield and the fluorescence lifetime. The decay from T$_{1}$ to S$_{0}$ is neglected in the model because it is much slower compared to other processes especially for the nanosecond optical pumping.}
     \label{Sfig:Jablonski diagram}
\end{figure}
\noindent In order to simulate the quantum amplification process, the simplified light propagation theory\cite{quan2021development,quan2023general}, based on the model shown in Supplementary Fig.~\ref{Sfig:Jablonski diagram} and the derived coupled differential equations (Supplementary Eqs.~3-7), was employed to quantitatively analyze the relationship between the triplet spin densities and the penetration depth $z$ under a certain pump intensity at the short time scale $t$ of applied light pulses.
\begin{equation}
\frac{\partial N_\text{T}(z,t)}{\partial t} = \frac{\Phi_{\text{ISC}}}{\tau_\text{F}} N_{10}(z,t) \
\end{equation}
\begin{equation}
\begin{split}
\frac{\partial N_0(z,t)}{\partial t} =&-\frac{c}{n_{\alpha}} \sigma_{\alpha} N_{\text{ph}\alpha}(z,t) N_0(z,t) - \frac{c}{n_{\gamma}} \sigma_{\gamma} N_{\text{ph}\gamma}(z,t) N_0(z,t) + \\
&\left(\frac{c}{n_{\alpha}} \sigma_{\alpha} N_{\text{ph}\alpha}(z,t) + \frac{c}{n_{\gamma}} \sigma_{\gamma} N_{\text{ph}\gamma}(z,t) + \frac{1 - \Phi_{\text{ISC}}}{\tau_\text{F}}\right) N_{10}(z,t)\
\end{split}
\end{equation}
\begin{equation}
\begin{split}
\frac{\partial N_{10}(z,t)}{\partial t} =& \frac{c}{n_{\alpha}} \sigma_{\alpha} N_{\text{ph}\alpha}(z,t) N_0(z,t) + \frac{c}{n_{\gamma}} \sigma_{\gamma} N_{\text{ph}\gamma}(z,t) N_0(z,t) - \\
&\left(\frac{c}{n_{\alpha}} \sigma_{\alpha} N_{\text{ph}\alpha}(z,t) + \frac{c}{n_{\gamma}} \sigma_{\gamma} N_{\text{ph}\gamma}(z,t) + \frac{1}{\tau_\text{F}}\right) N_{10}(z,t)\
\end{split}
\end{equation}
\begin{equation}
\frac{\partial N_{\text{ph}\alpha}(z,t)}{\partial t} + \frac{c}{n_{\alpha}} \frac{\partial N_{\text{ph}\alpha}(z,t)}{\partial z} = -\frac{c}{n_{\alpha}} \sigma_{\alpha} N_{\text{ph}\alpha}(z,t) (N_0(z,t) - N_{10}(z,t))\
\end{equation}
\begin{equation}
\frac{\partial N_{\text{ph}\gamma}(z,t)}{\partial t} + \frac{c}{n_{\gamma}} \frac{\partial N_{\text{ph}\gamma}(z,t)}{\partial z} = -\frac{c}{n_{\gamma}} \sigma_{\gamma} N_{\text{ph}\gamma}(z,t) (N_0(z,t) - N_{10}(z,t))\
\end{equation}
\noindent $N_\textrm{T}$, $N_\textrm{0}$ and $N_\textrm{10}$ are the spin densities in the triplet state (i.e. T$_\textrm{1}$), the singlet ground state and the singlet excited state. Since the crystal is birefringent, the laser beam is split into two rays - a fast ray and a slow ray, with the associated photon densities \(N_{\text{ph}\alpha}\) and \(N_{\text{ph}\gamma}\). The corresponded refractive indices are \(n_{\alpha}\) and \(n_{\gamma}\)\cite{sundararajan1936optical}, and the absorption cross sections are $\sigma_\alpha$ and $\sigma_\gamma$, respectively. $c$ is the speed of light in vacuum. $\Phi_\textrm{ISC}$ is the ISC yield\cite{takeda2002zero} and $\tau_\textrm{F}$ is the fluorescence lifetime\cite{patterson1984intersystem} for the pentacene molecules doped in \textit{p}-terphenyl. The values and expressions of the above quantities are summarized in Supplementary Table.~\ref{tab:Relevant parameters}. 
\FloatBarrier 
\begin{table}[htbp]
    \centering
    \begin{tabular}{ccccccc}
        \hline
          Parameter & $n_{\alpha}$ & $n_{\gamma}$ & $\Phi_{\text{ISC}}$ & $\tau_\text{F}$&$\frac{{\mathrm{\alpha_{px}}\omega}}{{4\pi \varepsilon_{0} c}}$&$\frac{{\mathrm{\alpha_{py}} \omega}}{{4 \pi \varepsilon_{0} c}}$ \\
        \hline
        Value &1.584 &2.004 &62.5\% &9 ns &$0.53 \times 10^{-18} \, \mathrm{cm}^{2}$&$0.29 \times 10^{-18} \, \mathrm{cm}^{2} $\\
        \hline
    \end{tabular}\\
     \vspace{1em}  
    \begin{tabular}{ccccc}
        \hline
          Parameter & $\sigma_{\alpha}$ & $\sigma_{\gamma}$ \\
        \hline
        Expression &$\frac{1}{9 \varepsilon_{0}} \left[ \frac{(n_{\alpha}^2 + 2)^2}{n_{\alpha}} \right] \mathrm{\alpha_{py}} \left[ \sin^2(\varphi) \right] \frac{\omega}{c} $&$\frac{1}{9 \varepsilon_{0}} \left[ \frac{(n_{\gamma}^2 + 2)^2}{n_{\gamma}} \right] \mathrm{\alpha_{px}} \frac{\omega}{c} $\\
        \hline
    \end{tabular}
    \caption{Parameter values and expressions employed in the simulation}
    \label{tab:Relevant parameters}
\end{table}
\noindent Where $\varepsilon_{0}$ is the permittivity of free space, $\omega$ is the angular frequency of the pump light, $\varphi$ is the angle between the b-axis and the Y$_1$, which is approximately 33°. $\mathrm{\alpha_{px}}$ and $\mathrm{\alpha_{py}}$ are the diagonal elements of the molecular polarizability tensor of pentacene\cite{quan2023general}. By inserting the values of the parameters in the above coupled differential equations for the numerical calculations via Mathematica, the total number of the triplet spins (the sum of the spin populations in T$_\textrm{-1}$, T$_\textrm{0}$ and T$_\textrm{+1}$) generated by the optical pumped can be obtained by integrating $N_\textrm{T}(z,t)$ over the penetration depth and then multiplying with the laser pumped area of the crystal (1 mm$\times$2 mm). Apparently, the laser pumped area can be improved in the future for gaining more triplet spins.

\printbibliography